\documentclass[conference,12pt,onecolumn]{IEEEtran}

\usepackage[justification=centering]{caption}

\ifCLASSINFOpdf
   \usepackage[pdftex]{graphicx}
   \graphicspath{{./fig/}}
   \DeclareGraphicsExtensions{.png}
\else
\fi
\usepackage{url}

\usepackage{subfigure}

\usepackage[utf8]{inputenc}
\begin{document}
%
\title{Using Fault Injection on the Nanosatellite Subsystems Integration Testing}

\author{\IEEEauthorblockN{Carlos L. G. Batista \IEEEauthorrefmark{1},
André Corsetti \IEEEauthorrefmark{2},
Maria de Fátima Mattiello-Francisco \IEEEauthorrefmark{1}}\\
\IEEEauthorblockA{\IEEEauthorrefmark{1} National Institute for Space Research\\
São José dos Campos, Brazil\\
carlos.gomes@crn.inpe.br\\
fatima.mattiello@inpe.br\\
\\
\IEEEauthorrefmark{2} Omega7 Systems\\
São José dos Campos, Brazil\\
andrecor7@gmail.com}
}

%


\maketitle

\begin{abstract}
Since the 2000's, an increased number of nanosatellites have accessed space. However, studies show that  
the number of unsuccessful nanosatellite missions is very expressive. 
Moreover, these statistics are correlated to poor verification and validation processes used by hobbyists satellite developers because major space agencies keep high successful ratings even with small/nano satellites missions due to its rigorous V\&V processes.
Aiming to improve payloads integration testing of NanosatC-BR-2, a 2-U Cubesat based nanosatellite under development by INPE, the fault injection technique has been used.
It is very useful technique to test systems prototypes. This paper presents the design and implementation of a Failure Emulator Mechanism  (FEM) on $I^2C$ communication bus for testing the interaction among the NanosatC-BR2 subsystems, supporting  interoperability and robustness requirements verification.
The FEM is modelled to work at the communication bus emulating eventual faults of the communicating subsystems in the messages exchanged.
Using an Arduino board for the FEM and NI LabView environment it is possible to program the mechanism to inject different faults at the $I^2C$ bus during different operation modes.
Based on a serial architecture, the FEM will be able to intercept all messages and implement different faults as service (i.e. message lost, bit flip and other single upset events) and timing (i.e. delay) faults.
The FEM interface with the tester is designed in LabView environment. Control and observation facilities are available to generate and upload the faultload script to FEM Arduino board.
The proposed FEM architecture and its implementation are validated using two subsystems under testing prototypes:  the OnBoard  Data Handling Computer and the Langmuir Probe NanosatC-BR2 payload. For this analysis purpose, the prototypes simulate in two different Arduinos boards the expected behavior of each subsystem in the communication.
\end{abstract}


\IEEEpeerreviewmaketitle

\section{Introduction}

In the last 20 years even more nanosatellites have been launched.
In special, the CubeSat Standard, developed at the earlies 2000', became a great option to access the space at low cost.

According to an estimation  at  Nanosatellite \& CubeSat Database \cite{nano2017database}, more than 550 nanosats have been launched until January 2017.

\begin{figure}
    \centering
    \includegraphics[width=0.5\textwidth]{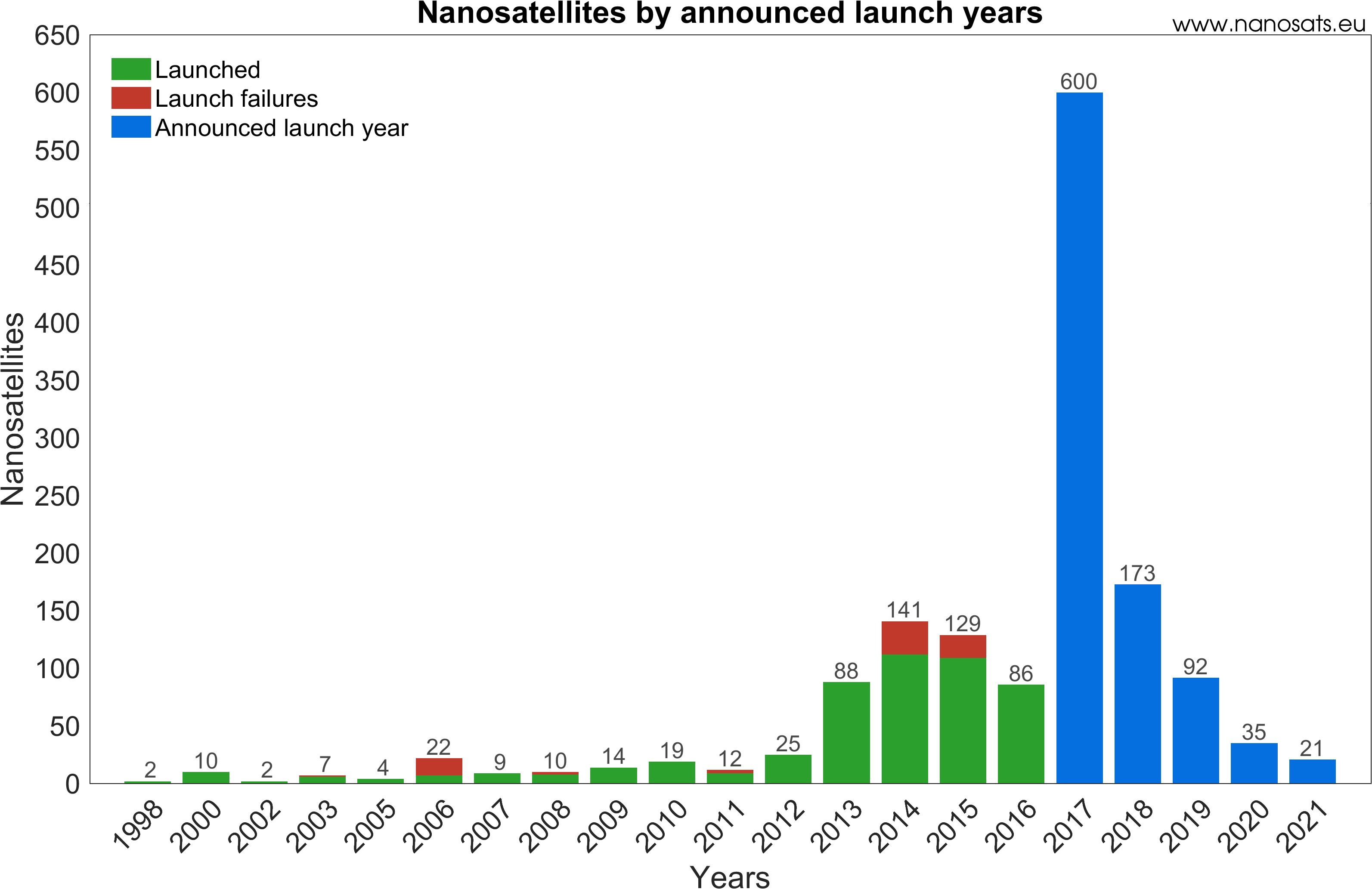}
    \caption{Nanosatellites launches over the years until 2021 (www.nanosats.eu)}
    \label{fig:nano_database}
\end{figure}

The Figure \ref{fig:nano_database} shows the announced nanosat launches over the years until 2021.
What we could notice from that data is an increase number of nanosat based missions all over the world during the past years.
Recent studies indicate a 50\% successful rate of these nanosat missions in general with better rates (87\% of success) among the major space agencies and traditional companies.  However among minor/hobbyists developers  the successfull rates reach around 40\% only\cite{swartwout2016secondary}.
What these rates means? The general consensus associates this to a lack of good Verification \& Validation processes in the development of these small space projects.

Rigorous V\&V practices commonly consume a great amount of time and money,
which comes on the opposite direction of the nanosat/CubeSat Mission, usually cheap and short development cycle.

In Brazil, the nanosat/CubeSat boom was not different.
During the past 10 years a great number of CubeSat Projects started (i.e. CONASAT, Serpens, AESP-14, UbatubaSat, NanosatC-BR).

Focusing on this emergent technology and aiming at contributing with a better V\&V processes  for these missions, this paper addresses the use of fault injection technique. 

A Failure Emulator Mechanism (FEM) is proposed to inject faults in the communication channel  I2C in order to support  NanoSat integration tests. The paper is structured as following. The section II describes briefly both the fault injection concept and the Test System where the proposed FEM shall act. Section III presents the FEM functions based on failure models and FEM architecture. Section IV describes the validation of FEM implementation using NanosaC-Br2 as case study. Finally, section V concludes the work.

\section{Concepts}

\subsection{Fault Injection}

Over the years  Fault Injection has been known as an effective test execution technique supporting  V\&V methodologies and processes.
The approach consists on deliberate inserting faults into a system in a way that emulates faults present in the system \cite{arlat1989fault}.

For this point and further, it will be considered the following concepts: a fault causes an error (unexpected internal state) that leads to a failure (crash of the system or possible degradation of the services) as defined by \cite{avizienis2004basic}.

Faults could be divided at two different categories, \textit{software} and \textit{hardware faults}.
Hardware faults, for example a bit flip caused by excessive radiation exposure, are difficult to perform in labs and, normally, destructive. Which are unavailable when we consider the low resources in a CubeSat mission. Software faults implies in low cost for implementation once the fault can be modeled and simulated with, \textit{a priori}, few modifications at the core system. But at the final stages of a space system project modifications are not allowed and this kind of test, injecting faults at software level, is unable to perform.

\subsection{Failure Emulator Mechanism}

The FEM concept adopted in this work follows  early studies \cite{mattiello2009inrob,mattiello2012inrob} .
FEM is a test execution mechanism that can inject faults into the message exchanged between two software intensive subsystems at the communication channel.
Thus FEM emulates failures in the communication channel (i.e. a bit-flip in a register could be emulated as a bit-flip at a sent package).

The goal of a FEM is to act in the communication channel intercepting the exchange messages and inserting (or not) faults that emulates failures of the Systems Under Test (SUT).

The FEM avoids no direct modifications in the software implementation enabling to emulate hardware failures from controlled communication faults.
As this, we can measure dependability characteristics of communicating subsystems and verify  their interoperability  and robustness requirements at the final stages of integration tests with hardware in the loop.

\subsection{Proposed Test System}

\begin{figure}
    \centering
    \includegraphics[width=0.5\textwidth]{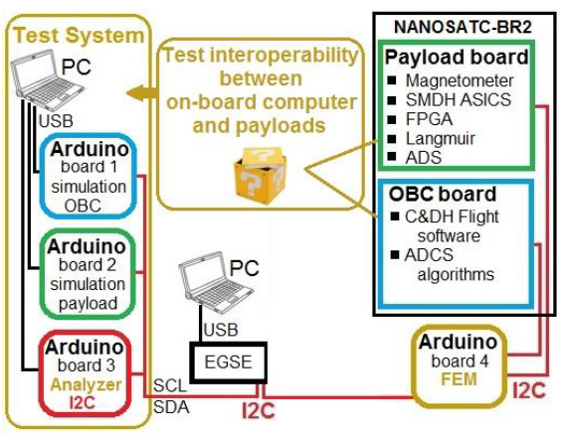}
    \caption{Test System on Scenario including the FEM \cite{conceicao2016dependability}}
    \label{fig:carlos_test_sys}
\end{figure}

The FEM mechanism shall be part ofa Test System.  In order to support the integration testing of Nanosatellite subsystems, the  FEM shall be part of the Test System  presented in Figure  \ref{fig:carlos_test_sys} . This Test System was designed to support  NanosatC-BR2  (a 2U Cubesat) integration testing  also under developed at the National Institute for Space Research (INPE) at São José dos Campos, Brazil.
The Test System  \ref{fig:carlos_test_sys} comprises Arduino boards connected by means of I2C bus, as extension of Nanosatellite regular bus. 

Thus the Test System can support the interoperability verification between the OnBoard \& Data Handling Computer -- OBC -- and the Payloads -- PLD –- on board  NanoSatC-BR2.

All tests are supported by an the EGSE -- Eletric Ground Support Equipment -- used to provide electric power and electronic measure points for the satellite when it is still on the ground.

One can observe that the FEM is also implemented by an Arduino that acts in the proposed Test System as a new element in the   communication channel  (I2C bus) between the OBC and PLD, referred as System Under Testing (SUT)
The objective of the FEM  is to support the integration test of each two SUTs at abnormal controlled conditions and to ensure that erroneous messages can be generated as required by the test case being run \cite{martins2003tool}.

The NanoSatC-BR2 project has five PLD planned to be on board: a magnetotorquer, an atitude determination subsystem, a Langmuir Probe, a SMDH ASICS and a radiation resistent FPGA, all payloads developed by institutes and universities from Brazil.

\section{Development}

\subsection{Model}

Based on an initial model \cite{weller2015inrob} that exemplifies the behavior of a generic FEM it is possible to derive an initial and feasible model for a nanosat fault injector.
Bonfiglio \cite{bonfiglio2015software} has modeled the failures of a system as failures of its services and classified them as: value (i.e. incorrect value), provision (i.e. output anything) and time (i.e. delayed message).

\begin{figure}[h!]
    \centering
    \includegraphics[width=0.5\textwidth]{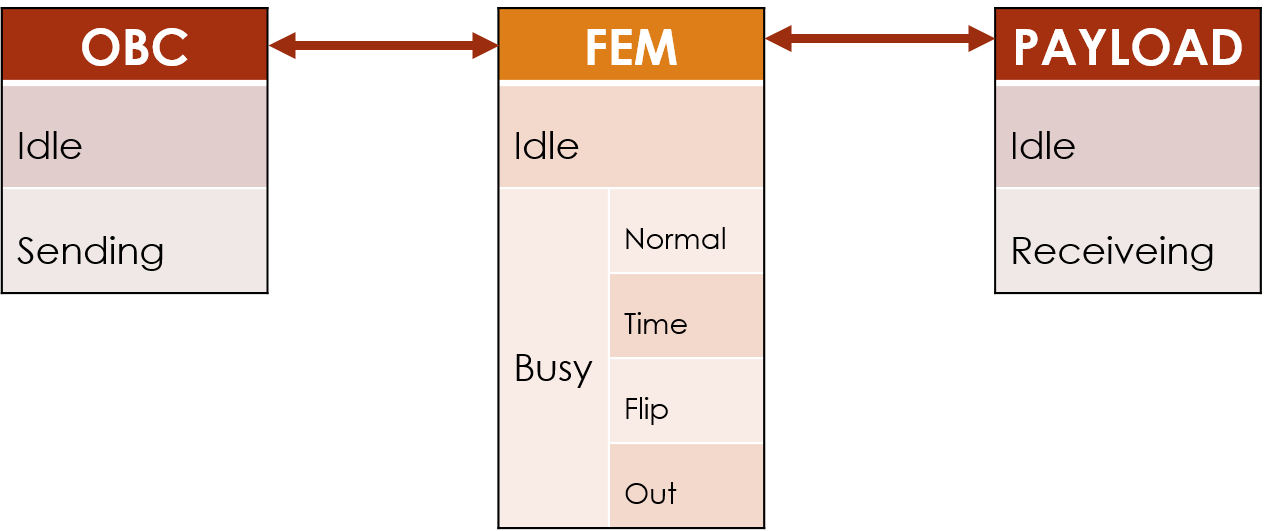}
    \caption{Class Diagram for FEM, OBC and PLD}
    \label{fig:fem_model}
\end{figure}

The Figure \ref{fig:fem_model} shows the class diagram with the initial model used for the FEM and simple behavioral models for OBC and PLD.

One can observe  that the FEM model has two different major modes: Idle and Busy.
The Idle mode represents no action at the messages traffic between the OBC and the PLD.
In the Busy mode, FEM starts to interact with the communication channel: monitoring (Normal), bit-fliping (Flip), timeout delaying (Delay) and changing (Out) the messages.
    
\begin{figure}[t!]
    \centering
    \caption{Simulated Faults at OBC-FEM-PLD System}
    
    \subfigure[Normal]{\includegraphics[width=0.2\textwidth]{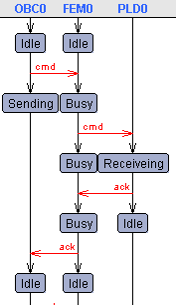}\label{fig:errors-n}}
    \subfigure[Timeout Delay]{\includegraphics[width=0.2\textwidth]{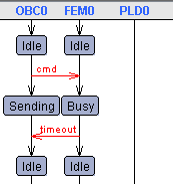}\label{fig:errors-t}}
    \subfigure[Bit-Flip]{\includegraphics[width=0.2\textwidth]{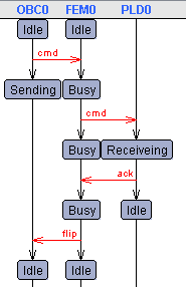}\label{fig:errors-f}}
    \subfigure[Out of Range Value]{\includegraphics[width=0.2\textwidth]{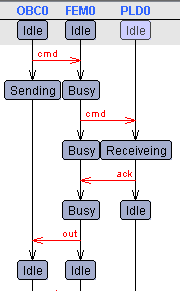}\label{fig:errors-o}}
    
    \label{fig:errors}
\end{figure}

This behavior has been demonstrated using a model-checking tool, UPPAAL, resulting at the Figure \ref{fig:errors} which consists of the expected behavior for each one of the faults injected by the FEM at the OBC-PLD comunication. On Subfigures: \ref{fig:errors-n}, the normal functioning, \ref{fig:errors-t}, the delaying of a OBC request until a timeout failure, \ref{fig:errors-f}, the bit-fliping of a PLD acknowledge message and \ref{fig:errors-o}, the change of a PLD response to an unexpected value.

For the OBC and PLD models, only theirs behavior has been implemented.
The payload subsystem chosen for this study is the NanosatC-BR2 Langmuir's Probe (SLP) payload, an instrument for ionospheric plasma measurement, essentially an analog reader and an A/D Converter.

\begin{figure}[h]
    \centering
    \subfigure[OBC command to SLP to Start Analog Read]{\includegraphics[width=0.3\textwidth]{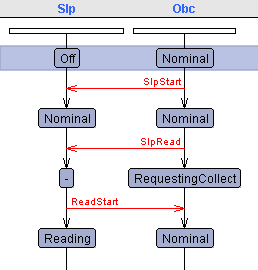}\label{fig:cmd}}
    \subfigure[OBC requesting Data from SLP]{\includegraphics[width=0.3\textwidth]{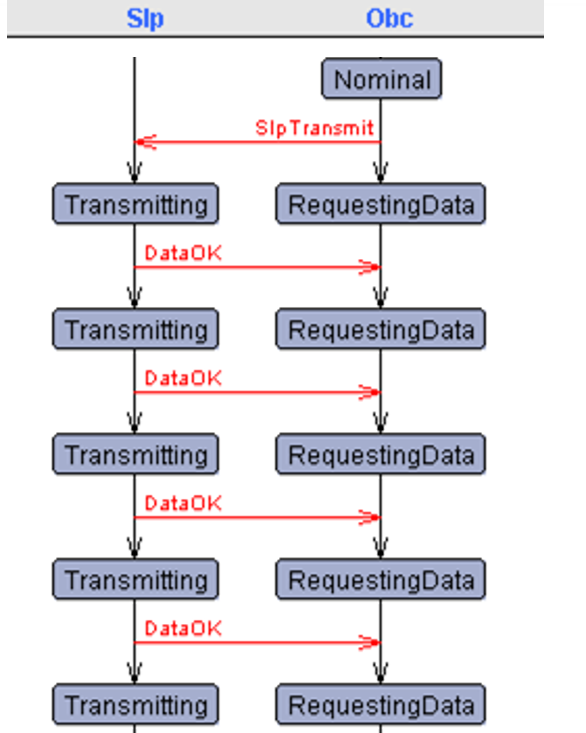}\label{fig:req}}
    \subfigure[OBC command to SLP to End the Transmission]{\includegraphics[width=0.3\textwidth]{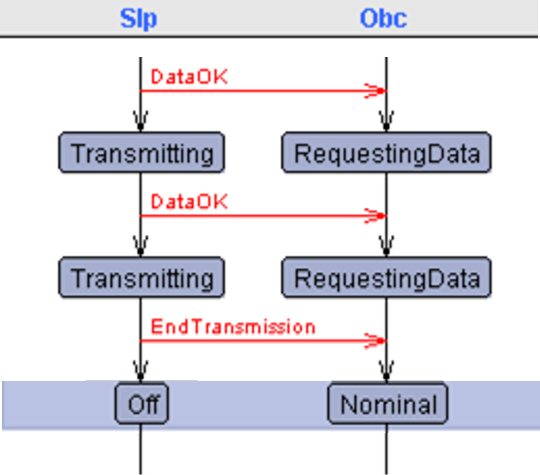}\label{fig:end}}
    \caption{Modeling the SUT behaviour.}
    \label{fig:behavior}
\end{figure}

The interface between the two subsystems is simple as shown in Figure \ref{fig:behavior} where the OBC commands the SLP to start the analog read, then the OBC request the data acquired from SLP and at last the OBC commands the SLP to end its transmission and to return to an off mode.

\subsection{Architecture}

\begin{figure}[ht]
    \centering
    \includegraphics[width=0.35\textwidth]{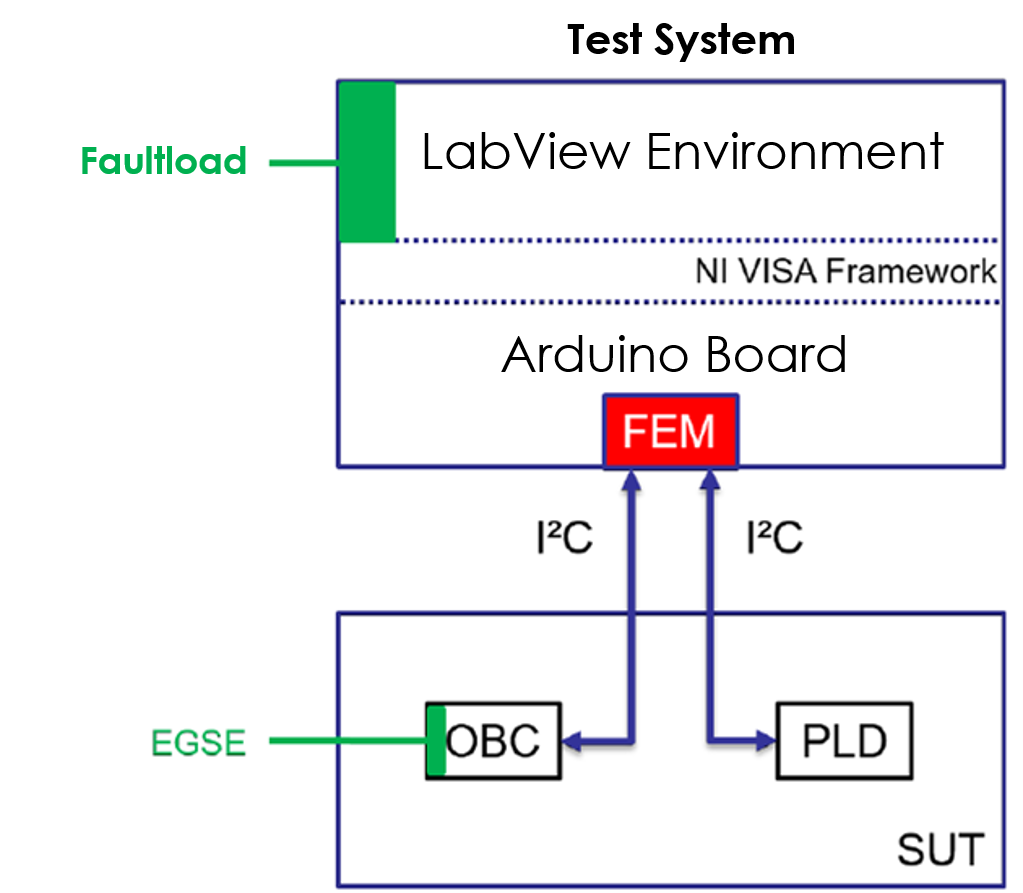}
    \caption{FEM Architecture}
    \label{fig:fem_arch}
\end{figure}

The FEM architecture is aligned with the CubeSat standard development philosophy, which is followed by the most of the Brazilian nanosat projects.
These projects adopted the 104-pin CubeSat Kit Bus, a connector standard for a sort of \textit{plug and play} stack philosophy.
On this bus, power and communication works in a parallel way where all subsystems on board the satellite use a common power feed bus (all have contact with the same power buses) and they also \textit{listen} at the same communication bus.
So, for communication, these projects chose the I2C protocol, a simple standard based on Master-asks-Slave-responds communication without any kind of error detection or error correction methods.
Therefore, on I2C, all the slave devices can hear the communication \cite{philips2003manual}, the FEM should be able to intercept and modify (or not), as defined by the model, the messages exchanged on the bus.
The Figure \ref{fig:fem_arch} exemplifies the idea of a serial FEM, the OBC is the Master device and the PLD works as the Slave device, which is able to intercept the communication and inject faults from a previous loaded \textit{faultload}.
All the implementation of the FEM model and its modes has been done on an Arduino board as shown on the architecture.
An Arduino Uno was used counting on\verb|<Wire.h>|, \verb|<TWI.h>| and \verb|<SoftI2CMaster.h>| libraries at the Arduino IDE on a MacBook.

The library \verb|<SoftI2CMaster.h>| enables the use of different pins at the board to be used as \textit{Two Wire Interface} -- TWI -- necessary for I2C SCL (clock) and SDA (data).
The Arduino board embeds  all FEM functions but the environment with the tester interface and the script need to be uploaded with the test cases.  Thus, the faultload  is part of  LabView application (National Instruments), as shown in Figure \ref{fig:fem_arch}

The application is responsible to create a file with the test cases (handmaded or automatic generated) which will be uploaded throught NI VISA Framework, a serial communication framework for LabView, to the Arduino board with three parameters:

\begin{itemize}
    \item Where? -- Inject a fault on Master-FEM side or on FEM-Slave Side.
    \item When? -- The exact message to inject a fault.
    \item What? -- The nature of the fault: time, provision or value.
\end{itemize}

\section{First Results}

The first results came from implementing all the models (FEM, OBC and SLP) on Arduino Uno boards and using the serial monitor from Arduino IDE to monitor the behavior. 

\begin{figure}[hb]
    \centering
    \subfigure[OBC Serial Monitor]{\includegraphics[width=0.25\textwidth]{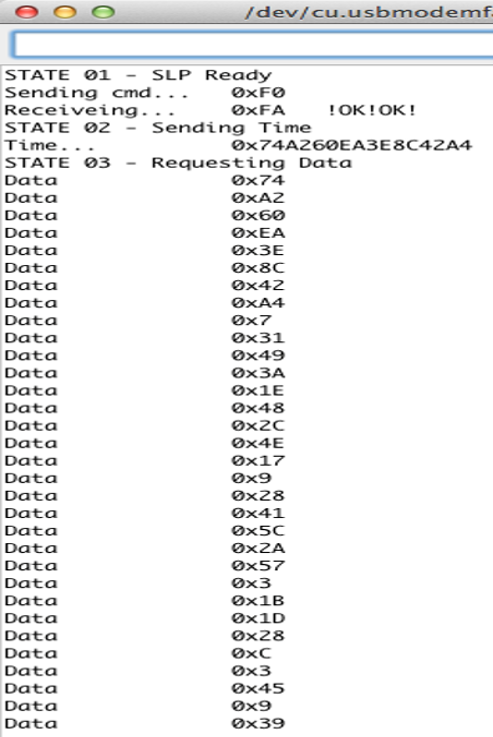}\label{fig:norm-obc}}
    \subfigure[SLP Serial Monitor]{\includegraphics[width=0.25\textwidth]{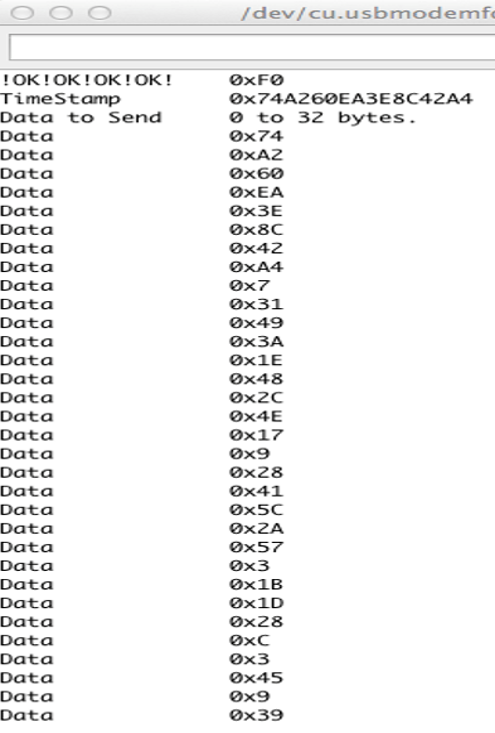}\label{fig:norm-slp}}
    \subfigure[FEM Serial Monitor]{\includegraphics[width=0.3\textwidth]{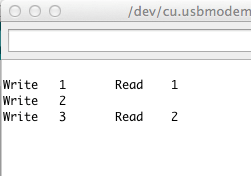}\label{fig:norm-fem}}
    \caption{Normal Operation of FEM}
    \label{fig:normal}
\end{figure}

The results related to the normal operation of FEM, where it just monitors the communication channel, are showed on Figure \ref{fig:normal}.
Pay attention that during this mode, the FEM \textit{listens} to the channel and counts (\ref{fig:norm-fem}) the number of Writes (requests from Master) and Reads (responses from Slaves) this allows the tester to determinate when the fault will be injected and observe the behavior of each subsystem (\ref{fig:norm-obc},\ref{fig:norm-slp}) for requirements verification.

When fault had been injected, the behavior of the I2C is still the same.
Once it does not have error detection, faults in time domain, for example, became just \verb|0xFF| reads on SDA bus.
The same expected read data in a complete failure of the SLP (Slave), because the SDA is left on \verb|HIGH| by the OBC (Master).

The other domains of faults, such as values out of range and bit-flips, depend on SUT to have implementations for error detection and correction, if needed.
Some tests had been done using an oscilloscope and its digital analyzer to check the delay that  FEM could possible inject at the communication and if this could also be a problem in future implementations.

\begin{figure}[ht]
    \centering
    \includegraphics[width=0.5\textwidth]{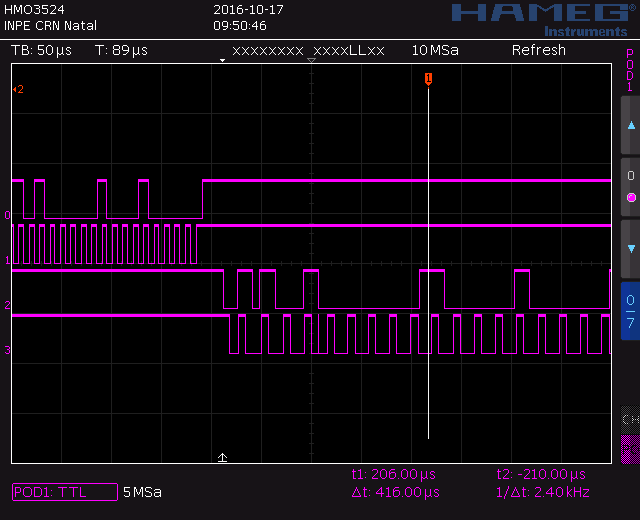}
    \caption{$I^2C$ Bus on Oscilloscope: Master Send}
    \label{fig:norm-osci}
\end{figure}

But due to a low time resolution, the results became inconclusive as shown at Figure \ref{fig:norm-osci}.
It is easy to see a huge delay on the communication, but the I2C itself seems to not detect directly these kind of time issue. Once more, it is an error to be detected by the SUT implementations themselves.

The upper two first traces belong to the OBC-FEM side. As consequence, the lower two traces belong to the FEM-SLP side.

Other results from first impressions at these tests is the lack of robustness of the I2C itself.
Random faults and errors are common at this interface, since from a simple pull-up resistors mismatch between the devices until a minimal logical voltage differences and no synchronisms between SDA and SCL buses.

\section{Conclusion}

Nanosat missions across the world and particularly in Brazil are proving they are a solid platform to reach the space at low cost and to flight new technologies for on orbit qualification purpose.
But it is necessary the use of best practices of development along the product life cycle, from conception to operation.  The challenge is how  the existing V\&V process can be downsized maintaining the nanosat and CubeSat philosophy, simple and fast.

The fault injection has already proved itself as an efficient tool for software requirements verification but its use on integration tests of space systems is still a step to be reached.

This paper tries to put a light at this approach and focus that with a good model driven design it is possible to reach the agility and quality level required for a good V\&V process on Nanosat/CubeSat mission even with low budget.
The results show that  the models and I2C are exciting even with the problems on the interface itself because it is possible to create a controlled environment to fault injection. Also the results demonstrate that the I2C is not the best option for this kind of missions.

More implementations will come within this year in order to improve the FEM model and the NI LabView environment as well. The Arduino libraries will be refined, the faults will be better implemented and FEM as a whole will be tested with real satellite subsystems to validate its use as part of a Test System in a Nanosat development process.


\section*{Acknowledgment}
The authors would like to thank CAPES, ETE/CSE INPE, CRN -- Natal, CRS -- Santa Maria.




\bibliographystyle{IEEEtran}
\bibliography{refe}
%



\end{document}